# HASCO: Towards Agile HArdware and Software CO-design for Tensor Computation

Qingcheng Xiao
*School of EECS*
*Peking University*
walkershaw@pku.edu.cn

Size Zheng
*School of EECS*
*Peking University*
zhengsz@pku.edu.cn

Bingzhe Wu
*School of EECS*
*Peking University*
wubingzhe@pku.edu.cn

Pengcheng Xu
*School of EECS*
*Peking University*
jsteward@pku.edu.cn

Xuehai Qian
*University of Southern California*
xuehai.qian@usc.edu

Yun Liang[†]
*School of EECS*
*Peking University*
ericlyun@pku.edu.cn

*Abstract*—Tensor computations overwhelm traditional general-purpose computing devices due to the large amounts of data and operations of the computations. They call for a holistic solution composed of both hardware acceleration and software mapping. Hardware/software (HW/SW) co-design optimizes the hardware and software in concert and produces high-quality solutions. There are two main challenges in the co-design flow. First, multiple methods exist to partition tensor computation and have different impacts on performance and energy efficiency. Besides, the hardware part must be implemented by the intrinsic functions of spatial accelerators. It is hard for programmers to identify and analyze the partitioning methods manually. Second, the overall design space composed of HW/SW partitioning, hardware optimization, and software optimization is huge. The design space needs to be efficiently explored.

To this end, we propose an agile co-design approach `HASCO` that provides an efficient HW/SW solution to dense tensor computation. We use tensor syntax trees as the unified IR, based on which we develop a two-step approach to identify partitioning methods. For each method, `HASCO` explores the hardware and software design spaces. We propose different algorithms for the explorations, as they have distinct objectives and evaluation costs. Concretely, we develop a multi-objective Bayesian optimization algorithm to explore hardware optimization. For software optimization, we use heuristic and Q-learning algorithms. Experiments demonstrate that `HASCO` achieves a 1.25X to 1.44X latency reduction through HW/SW co-design compared with developing the hardware and software separately.

## I. INTRODUCTION

Tensor computation is fundamental to many scientific and engineering applications, such as machine learning [4], [47], [67], [68], data mining [40], [53], [58], and quantum chemistry [16], [71]. Tensors are data organized in multi-dimensional arrays. Common tensor computations include matricized tensor times Khatri-Rao product (MTTKRP), tensor-times-matrix (TTM), general matrix multiply (GEMM), general matrix-vector multiplication (GEMV), generalized vector addition (AXPY), and convolution. More importantly, a real-world tensor application usually has multiple tensor computations, and each tensor computation can have multiple workloads. For instance, the semantic labeling application [38] includes dozens of GEMM and 2D convolution workloads that differ in tensor sizes.

For tensor applications, it is essential to develop a *holistic* solution that is a combination of *hardware acceleration* and *software mapping*. The conventional general-purpose processors suffer from the increasingly high complexity of tensor computation, which motivates specialized hardware acceleration. Recently, spatial accelerators implemented on FPGAs and ASICs have been shown to be efficient hardware architectures for tensor computation due to their massive parallelism and high energy efficiency [13], [14], [26], [27], [35], [42], [45], [63], [69]. For instance, Google Cloud TPU [35], [56], an ASIC processing neural networks, can reduce the training time by 27X at a 38% lower cost than NVIDIA V100 GPU clusters. On the other hand, the success of an end-to-end acceleration solution hinges largely on the software mapping or compilation. For tensor computation accelerators, the software mapping is responsible for splitting a large tensor into sub-tensors and invoking the corresponding hardware execution, as the accelerator can only handle a fixed size of tensor at a time. Software mapping is crucial for performance optimization. For instance, compared with manually calling tensor cores of the V100 GPUs, an optimized software CUTLASS [20] can achieve up to 1.73X performance improvement.

Though dedicated hardware and software optimizations have progressed considerably for tensor computation, they primarily focus on *either the hardware part [25], [36], [39], [62], [81] or the software part [12], [19], [32], [60], [85]*. Optimizing the two parts in isolation inevitably suffers from sub-optimal solutions confined in a local design space. While seemingly appealing, there has been less attention on hardware/software co-design for tensor computation [2], [7], [72]. This is largely because the design of hardware and software components influence each other, and the joint design space can be huge. A general approach to tackle the co-design problem is to develop a unified intermediate representation (IR), based on which designers can partition the hardware and software, optimize, and synthesize the hardware and software. However, developing such a general IR and synthesizing arbitrary hardware are challenging [30], [76].

[†]Corresponding Author

In this work, we provide a co-design approach specific to tensor computation. As tensor computation can be described using nested loops, we naturally embed loop information into our IR design for tensor computation partitioning, optimization, and implementation. A subset of the loops are implemented using spatial hardware accelerators, and the remaining loops are implemented using software programs. The fundamental questions are: 1) how to define the interface between the hardware accelerators and the software programs, 2) how to navigate the huge design space for each part.

The accelerator designed for tensor computation typically supports one or a set of specific functions, which are termed as *hardware intrinsics*. For instance, the hardware intrinsic of Gemmini accelerators [24] is a GEMM function. The intrinsics of NVDLA accelerators [21] include 2D convolutions, pooling, activation functions, etc. Accordingly, we term the HW/SW interface for tensor computation as *tensorize*, which determines how to divide the tensor computation into sub-workloads and map the sub-workloads onto hardware intrinsics. The challenge is that **there exist multiple tensorize choices, which have significant impacts on performance and energy efficiency**. For instance, we can choose different loop subsets of a 2D convolution to form GEMV sub-workloads. Also, we can divide the 2D convolution into dot product, 1D/2D convolutions, GEMM, and other sub-workloads, leading to numerous tensorize choices. The sub-workloads are likely to vary a lot in performance due to different levels of data locality, reuse opportunities, and padding styles. Thus, designers do not yet have a systematic approach to select from these choices.

The tensorize interface separates hardware and software so that each can be optimized separately. However, the hardware and software design spaces are still huge. Accelerator parameters consisting of bandwidth, memory, parallelism, dataflow, etc., determine the detailed implementation of the intrinsic functions. Collectively, **these parameters form a huge design space, which cannot be exhaustively searched**. For example, the legal design space of a GEMM accelerator [24] is on the order of $10^9$. Besides, developers need to prune the design space for different performance (latency/throughput), power, and area constraints. On the other hand, **software mapping for various tensor computations requires deep comprehension of the target accelerator**. For instance, loop reordering changes data locality, which impacts the efficiency of the memory hierarchy. The factors of loop splitting determine the size of a sub-workload, which is restricted by the hardware dataflow and on-chip memory size. Programmers call for efficient approaches to explore the design spaces.

In this paper, we propose HASCO, an agile co-design approach for tensor computation. HASCO jointly optimizes the hardware-software interface, hardware parameters, and software optimizations. First, we define tensor syntax trees with loop information and use them as a unified IR for tensor computation. The tensor syntax trees expose numerous tensorize choices, which require hardware acceleration with different hardware intrinsics. Given a hardware intrinsic, we explore different tensorize choices using a two-step matching approach. Synthesizing all the possible intrinsics into hardware is challenging and beyond the scope of this paper. In practice, we limit the hardware intrinsics to a subset of commonly used intrinsics (GEMV, GEMM, convolution, and dot product). Then, we generate holistic solutions for each tensorize choice and compare them. Concretely, we treat the hardware exploration as a multi-objective problem, where performance, power, and area are optimized. We develop a Bayesian optimization algorithm to find the Pareto set of hardware parameters. For the software, we use heuristic and Q-learning searching algorithms to find the optimized software mapping. The optimization for hardware and software are inherently correlated. The Bayesian-based hardware optimization uses the software latency as the performance metric, while the heuristic and Q-learning-based software optimization tailors the software mappings for the hardware parameters.

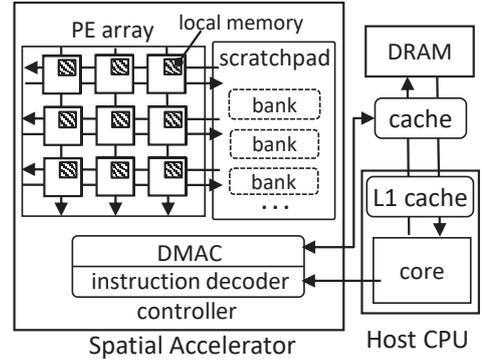

Fig. 1: A system overview of spatial accelerators.

HASCO mainly targets tensor applications with various tensor computations and workloads. Through co-design, the hardware part generates a specialized accelerator (hardware intrinsic) shared by all the tensor computations of the application. The software part retains flexibility by providing different optimized software mapping onto the accelerator for each workload. To sum up, our key contributions include:

- We propose HASCO to co-design hardware accelerators and software mapping in concert. HASCO offers a holistic solution to tensor computations.
- We propose efficient algorithms to explore the hardware-software interface (tensorize).
- We develop heuristic, Q-learning, and Bayesian optimization algorithms to explore the design spaces efficiently.

The source code of HASCO is publicly available at Github (https://github.com/pku-liang/HASCO). Experiments demonstrate that HASCO achieves a 1.25X to 1.44X latency reduction through HW/SW co-design compared with developing the hardware and software separately. HASCO speeds up the hardware design space exploration by 2.5X and achieves a 1.19X hypervolume compared with NSGAII. HASCO also optimizes the software by 3.17X and 1.21X compared to a library implementation and AutoTVM, respectively.

## II. BACKGROUND AND MOTIVATION

### A. Spatial Accelerators

Spatial accelerators [25], [35], [62], [81], [82] have been successfully employed to accelerate tensor computation. They expose low-level data transfer and computation through ISAs and can support multiple dataflows, which are the ways tensors are distributed and reused [41]. HASCO generates the spatial accelerators shown in Figure 1, which consist of three basic components: a 1D/2D array of processing elements (PEs), a memory system, and a controller. In the PE array, dozens of PEs communicate via on-chip interconnections between them and enable massive parallelism. Each PE performs computations with ALUs and registers. The memory system consists of a scratchpad memory shared by all the PEs and optional local memories within PEs. The scratchpad memory can be partitioned into banks to support concurrent data accesses. In the controller, there is an instruction decoder and a direct memory access controller (DMAC). The instruction decoder fetches and decodes instructions controlling the PEs and the DMAC. The DMAC moves large chunks of data between the DRAM and the accelerator's scratchpad through the off-chip caches. The data movement between the scratchpad and PEs is controlled by load/store instructions.

In this paper, we target the spatial accelerators with memory and control systems specified in Figure 1. Commercial tensor accelerators could have more complex memory and control systems. For instance, Google TPU uses dedicated buffers for weights and results and designs sophisticated synchronization. We leave the support for such complex designs to future work.

### B. HW/SW Interface: Tensorize

Here, we use an example of mapping 2D convolutions to accelerators with GEMM hardware intrinsics. In Listing 1, *Conv_workload_1* and *Conv_workload_2* are two convolution workloads (layers) from ResNet [28], and *GEMM_intrin* is the GEMM hardware intrinsic. A 2D convolution can be expressed as $C[k, x, y] = \sum A[c, x+r, y+s] * B[k, c, r, s]$, where tensor B is filters, A is input feature maps, and C is output feature maps. In *Conv_workload_1*, A convolves B with size $64 \times 64 \times 3 \times 3$ to produce C with size $64 \times 56 \times 56$. Similarly, the GEMM intrinsic is expressed as $L[i, j] = \sum M[i, k] * N[k, j]$.

The GEMM intrinsic can only process GEMM computations with a fixed size ($16 \times 16$ here), which is determined by the PE array shape of the accelerator. Directly calling the intrinsic function from the host CPU is inefficient, as the data movement between the host and the accelerator would be frequent and short. Hence, we need HW/SW interfaces to tensorize computations, transfer data in bursts, and invoke the intrinsic multiple times. In Listing 1, *Tensorized_GEMM_1* and *Tensorized_GEMM_2* are the interfaces for *Conv_workload_1* and *Conv_workload_2*, respectively. They process fixed but larger GEMM sub-workloads compared with the intrinsic. The sub-workload sizes are mainly constrained by the scratchpad size and the burst length of the accelerator.

As we can express a sub-workload as the sub-loops of tensor computation, we achieve tensorization by loop splitting and reordering. In *Conv_workload_1*, computation in the k, x, and c loops form tensorized sub-workloads represented by *Tensorized_GEMM_1*. The k loop is split into two sub-loops represented by k1 and k2. Similar splitting is applied to the x and c loops. After reordering the loops, k2, x2, and c2 determine the size of the tensorized sub-workloads. The outer six loops (Line 2-7) form the software program, which launches the sub-workloads.

**Listing 1** Mapping 2D convolutions to GEMM accelerators.
```
1  def Conv_workload_1(A, B, C, ...):
2    for y in range(0, 56):
3      for r in range(0, 3):
4        for s in range(0, 3):                    Software
5          for k1 in range(0, 64, 32):            Program
6            for x1 in range(0, 56, 32):
7              for c1 in range(0, 64, 8):         HW/SW
8                Tensorized_GEMM_1(A, B, C, ...)  Partitioning
9
10 def Tensorized_GEMM_1(A, B, C, ...):
11   Tensor sA, sB, sC
12   sA = A[c1:c1+8, x1+r:x1+r+32, y+s]
13   sB = B[k1:k1+32, c1:c1+8, r, s]
14   for k2 in range(0, 32, 16):
15     for x2 in range(0, 32, 16):                Hardware
16       for c2 in range(0, 8):                   Accel.
17         M = sA[c2, x2:x2+16]
18         N = sB[k2:k2+16, c2]
19         L = GEMM_intrin(M, N, ...)             Intrinsic
20         sC[k2:k2+16, x2:x2+16] += L
21   C[k1:k1+32, x1:x1+32, y] += sC
22
23 def Conv_workload_2(A, B, C, ...):
24   ... Tensorized_GEMM_2(A, B, C, ...)
25
26 def Tensorized_GEMM_2(A, B, C, ...):
27   ... L = GEMM_intrin(M, N, ...) ...
```

In *Tensorized_GEMM_1*, $sA$, $sB$, and $sC$ are buffers in the scratchpad memory. The interface loads a subset of A and B into the scratchpad memory, performs GEMM computation, and then stores the result of C back to the DRAM. Specifically, it distributes the scratchpad buffers to PEs' local memories or registers ($M$, $N$, and $L$) and calls the intrinsic (*GEMM_intrin*) 32 times. The k2, x2, and c2 loops determine how the data are organized and computed. Their order needs to match the accelerator's dataflow, and their strides must be identical to the PE array shape. As will be introduced in Section VI-C, each interface is a sequence of compute and data movement instructions executed by the hardware accelerators. In *Tensorized_GEMM_1*, Lines 12, 13, and 21 are load/store instructions, and *GEMM_intrin* represents compute instructions. In short, tensorize interfaces are highly architecture-specific and must be carefully programmed. Our co-design flow can automatically infer it once the tensorize choice is made.

### C. Motivational Case Study

The hardware parameters and software optimizations are hard to determine but vital to performance. In this case study, we prototype two GEMM accelerators (GA_L and GA_S) with different parameters on FPGAs. GA_L has a $16 \times 16$ PE array and a 256 KB scratchpad memory, while GA_S has an $8 \times 8$ PE array and a 128 KB scratchpad. We run a set of optimized programs on both architectures.

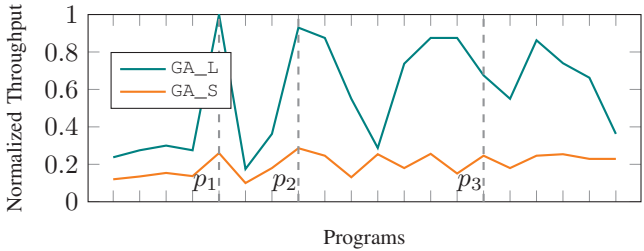

Fig. 2: Normalized throughput when running optimized programs on the two GEMM accelerators.

Figure 2 gives the results. The X-axis represents different software programs, and the Y-axis is the throughput normalized by the peak throughput of GA_L. First, we find that **software optimizations have a huge but unpredictable impact on the final performance**. We highlight three programs, p1 to p3, in the figure. Programs p1 and p2 have the same amount of on-chip computation but different loop orders. Program p3 has the same loop order as p1 but more on-chip computation. As shown, p1 achieves the peak performance for GA_L, which means loop orders (p2 v.s. p1) and tensorization all matter, and more on-chip computation does not necessarily result in higher performance (p3 v.s. p1). Also, p2 instead of p1 achieves the peak performance on GA_S, which suggests different hardware accelerators prefer different software optimizations. Second, **the design space exhibits complex trade-offs**. GA_L has a 4X larger PE array and a 2X larger scratchpad memory than GA_S. For our FPGA prototypes, GA_L consumes 2.58X more area and 1.49X more power and achieves 4.27X peak throughput improvement (122.33 MOPS v.s. 28.68 MOPS) compared with GA_S. Floor-planned ASIC designs also demonstrate a complex relation [24]. It is hard to choose accelerator parameters to meet the constraints of different scenarios. Also, this example only involves one hardware intrinsic and one workload. Programmers would face greater challenges given various tensor computations and intrinsics. To this end, we propose HASCO to explore the hardware and software design spaces in concert.

## III. HASCO

Figure 3 presents the workflow of HASCO. Users specify the computation workloads in a tensor application, the hardware generation method, and constraints in the input description. HASCO co-designs and outputs solutions for the application. A holistic solution consists of an accelerator shared by all the workloads within an application, hardware and software interfaces, and a software program per workload. We divide a co-design process into three steps, as shown in Figure 3:

**Step 1: HW/SW Partitioning.** HASCO first identifies tensorize choices representing HW/SW partitioning from tensor syntax trees. All these choices form the partition space, which will be explored with software design space in concert.

**Step 2: Solution Generation.** HASCO explores different hardware accelerators and software programs through design space exploration (DSE). We develop different DSE algorithms, as the hardware and software design spaces differ in optimization objectives and evaluation costs. The software DSE is usually performance-driven and can be fast if we fix the accelerator. The hardware DSE concerns multiple objectives like power and area in addition to performance. Each point in the hardware design space represents an accelerator instance. Evaluating design points may require prototyping accelerators, which is a lengthy and expensive process.

HASCO explores the hardware design space with a Multi-objective Bayesian Optimization (MOBO) algorithm and obtains the Pareto optimal accelerator parameters, as introduced in Section V-B. Based on these parameters, HASCO generates spatial accelerators with common intrinsics (GEMV, GEMM, convolution, and dot product) using off-the-shelf generators [21], [24], [33], [74], [76], [79] or our built-in Chisel generator. Then, HASCO explores software optimizations through heuristic and Q-learning algorithms, as detailed in Section VI. Also, HASCO automatically generates interfaces. A software program or an interface is specific to a workload and the accelerator.

**Step 3: Solution Tuning.** Performance metrics are collected by running the software on top of the accelerator. If the metrics violate the user constraints, they will drive the hardware DSE and generate a new accelerator. Accordingly, the software and interfaces are also re-generated. HASCO evaluates the metrics through mathematical models [41], [46], [59], [65], [66] and runtime profiling.

## IV. HW/SW PARTITIONING

### A. Tensorize Choices

We define *tensorize choices* as the ways to decompose a tensor computation into sub-workloads. The sub-workload sizes are determined by software. Take the GEMM computation (L = M $\times$ N) as an example. Figure 4 gives four tensorize choices. The first three choices form GEMV sub-workloads. Naturally, we can treat columns or rows of N as the vectors in GEMVs, as the #1 and #2 choices illustrate. However, choice #2 is illegal as it outputs incorrect results. Treating rows of M as the vectors like choice #3 is also legal if matrix transpositions are allowed. We can further multiply an element of M and a row of N to match AXPY, as choice #4 does. These choices differ in data padding, reuse, and locality. Many other tensorize choices exist for this simple example. All legal tensorize choices form the entire partition space. As the tensor dimensions increase, it is hard for programmers to identify and analyze different tensorize choices. HASCO provides an automatic approach to explore the partition space.

### B. Partition Space Generation

Here, we first discuss the tensorize choices for a specific hardware intrinsic. Given a hardware intrinsic and a tensor computation as inputs, we use a two-step approach to output all the legal tensorize choices that match the intrinsic. We define tensor syntax tree (TST) for both hardware intrinsic (intrinsic TST) and tensor computation (compute TST). TST that abstracts the loop and tensor information serves as the unified HW/SW IR for tensor computation. In a tensor syntax tree, each internal node is an operation (e.g., sum, add, multiply, and indexing), and the children of the node are the operands of the operator. An internal indexing node ([]) represents a tensor. Its leaf nodes are the loops accessing the tensor. Figure 5(b) illustrates the TSTs of the GEMM intrinsic and the 2D convolution. The intrinsic tree has four leaf nodes representing the four indexes in the notation $\sum M[i,k] * N[k,j]$. The compute tree has nine leaf nodes corresponding to the indexes in $\sum A[c, x+r, y+s] * B[k,c,r,s]$. The nodes $\mu_3$ and $\mu_6$ indicate the last dimension of A is accessed by the $y$ and $s$ loops. TSTs explicitly show the tensor dimensions and loops involved in an operation.

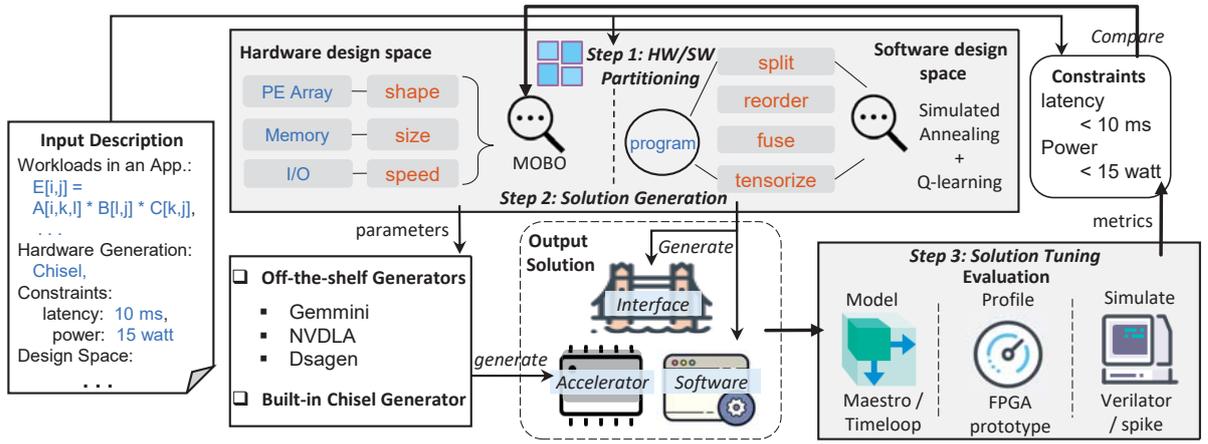

Fig. 3: The workflow of HASCO.

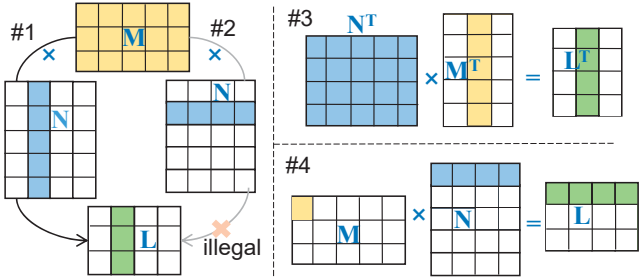

Fig. 4: Four tensorize choices for GEMM. The squares are data, and the colored ones form GEMVs or AXPY.

HASCO lowers both tensor computations and intrinsics into TSTs and performs a two-step approach: index matching and structure matching. ***In the index matching step***, HASCO enumerates the leaf nodes subsets of the compute tree. Given an intrinsic tree $Q$, a potential leaf subset $P$ must: ① have the same number of leaf nodes as $Q$ does, ② ensure a bijective mapping from each leaf node $\nu \in Q$ to a node $\mu \in P$. For instance, nodes $\nu_1, \nu_2 \in Q$ representing index $k$ in Figure 5(b). If $\nu_1 \leftrightarrow \mu_1$ and $\nu_2 \leftrightarrow \mu_2$, then $\mu_1, \mu_2$ must represent the same index ($c$ in this case). ***In the structure matching step***, HASCO finds the lowest common ancestors (LCAs) of every two nodes in the subset $P$ to match the internal nodes of the intrinsic tree $Q$. In the figure, node $\mu_4$ is the LCA of $\mu_3$ and $\mu_1$. If $\mu_1 \leftrightarrow \nu_1$ and $\mu_3 \leftrightarrow \nu_3$ are determined in the index matching step, we require $\mu_4$ to represent the same operation with the LCA of $\nu_1$ and $\nu_3$ (node $\nu_4$ in the figure). Another mapping $\mu_6 \leftrightarrow \nu_1$ that maps index $s$ to $k$ can also pass the index matching. However, node $\mu_5$, the LCA of $\mu_3$ and $\mu_6$, and $\mu_4$ represent different operations, leading to an illegal matching.

The two-step matching examines whether the hardware intrinsic can implement the sub-workload formed by the leaf subset. It does not restrict the order or range of the matched leaf nodes and allows more tensorize choices. For instance, the order of $\mu_1$ and $\mu_3$ in the compute tree differs from the order of $\nu_1$ and $\nu_3$ in the intrinsic tree. Besides, $\mu_1$ and $\mu_3$ are from non-adjacent dimensions (the first and last dimensions of A). Different node orders give different tensorize choices with data rearrangements, like the matrix transpositions of choice #3 in Figure 4. In addition, the matching does not decide the range of each node, such that the size of the sub-workload is flexible. Given a hardware intrinsic, the time complexity of the two-step matching is $O(C_m^n * l)$, where $n$ is the number of leaf nodes of the intrinsic TST, $m$ is the number of leaf nodes of the compute TST, and $l$ is the total number of nodes of the compute TST. In practice, $l$, $m$, and $n$ are often small ($m \leq n \leq 10$ and $l \leq 100$). For the case in Figure 5(b), the matching examines 126 leaf subsets and finds six legal tensorize choices in minutes.

If we allow any form of hardware intrinsic, for a compute TST with $m$ leaf nodes, its all $2^m - 1$ leaf nodes subsets are possible tensorize choices and form the entire partition space. For instance, the 2D convolution has 511 potential tensorize choices. However, it is infeasible to explore all the choices as each requires a specific hardware design, and the exploration will be extremely long. In practice, HASCO uses four common hardware intrinsics (GEMV, GEMM, convolution, and dot product) to decompose the workloads. As will be introduced in Section VI, the partition space of each intrinsic is included in the software design space. To make tensorize choices, HASCO generates holistic solutions for each choice and compares the performance metrics of the solutions.

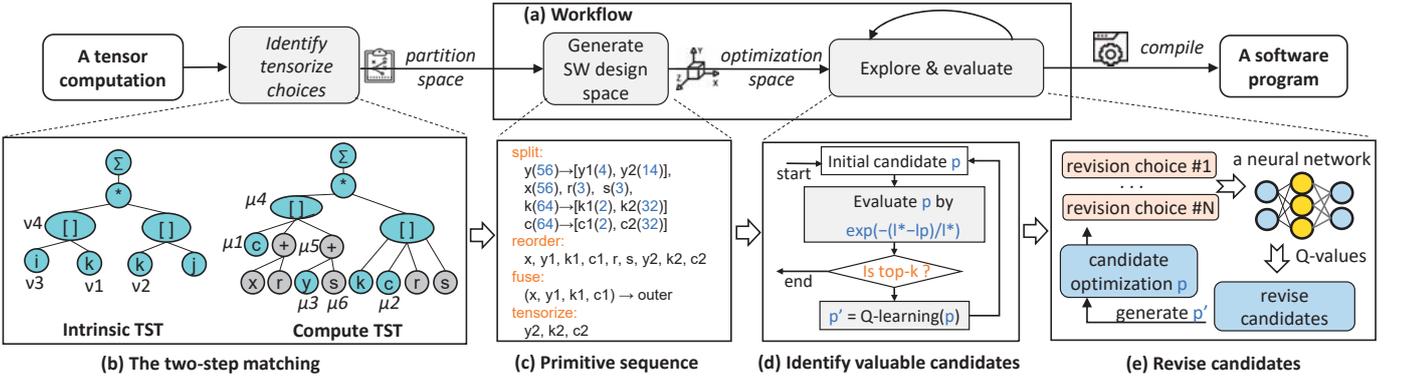

Fig. 5: Schedule a 2D convolution on GEMM accelerators. (a) Software optimization flow. (b) The two-step matching. (c) A primitive sequence representing an optimization for the convolution. (d) The heuristic algorithm. (e) The Q-learning algorithm.

## V. HARDWARE GENERATION

### A. Hardware Primitives and Design Space

We provide hardware primitives to define and prune hardware design space. As shown in Figure 6, the primitives describe three aspects of spatial accelerators: computation parallelism (*reshapeArray* and *linkPEs*), on-chip cache hierarchy (*addCache*, *distributeCache*, and *partitionBanks*), and off-chip memory access (*burstTransfer*). *reshapeArray* specifies the PE array shape and the intrinsic size. Parallel computation relies on the massive communications between PEs. We abstract common interconnection patterns and use *linkPEs* to specify them. Cache configurations (size, bank number, and distribution) also impact spatial accelerators' overall performance. Developers can use *addCache* to embed a scratchpad memory shared by all PEs. A scratchpad memory can be partitioned into multiple banks via *partitionBanks* to support concurrent accesses from PEs. It can be further distributed into each PE to form private local memories through *distributeCache*. Last, to speed up off-chip memory accesses, we can use *burstTransfer* to define a DMA controller between a cache and the DRAM.

We use a sequence of the parametric hardware primitives to form the skeleton of a spatial accelerator, and the primitive factors (accelerator parameters) compose the design space. Take the accelerator illustrated in Figure 1 as an example. Its design space is composed of the following parameters: [*scratchpad size*, *# scratchpad banks*, *local memory size*, *burst length of DMAC*, *maximal transfer size of DMAC*, *dataflow*, *PE array shape*]. Accordingly, the goal of hardware generation is to determine the factors of the hardware primitives. Listing 2 describes a systolic GEMM accelerator with the primitives. We first define a design space by specifying the hardware intrinsic (GEMM). The PE array is set as $16 \times 16$ and interconnected with a systolic pattern, where PEs receive data from their upstream neighbors and pass results downstream. The memory system is a 256 KB scratchpad without local memory. The DMAC bridges the scratchpad and the DRAM.

The hardware primitives only describe the architecture at a high level without specifying the underlying hardware implementation. From the primitives, HASCO uses generators to implement the real hardware. The off-the-shelf generators hide

**Listing 2** Describe a systolic GEMM accelerator.

```
1 acc = createArch(method = "Chisel", intrinsic =
    ↪ L[i, j]: M[i, k] * N[k , j])
2 # describe the PE array
3 acc.reshapeArray(16, 16)
4 acc.linkPEs("Systolic")
5 # describe the scratchpad and DMA
6 scratchpad = acc.addCache(256 * 1024)
7 acc.burstTransfer(scratchpad, 64, 128)
```

most architecture details from users and only expose a number of optimization knobs. HASCO can instantiate the generators with the determined primitive factors. In addition, we develop a Chisel [6] generator in HASCO, which translates the four common intrinsics (GEMV, GEMM, convolution, and dot product) and the hardware primitives into spatial accelerators.

### B. Accelerator Parameter Exploration

Once the design space is defined, HASCO starts to explore the accelerator parameters and optimize multiple performance metrics. As the correlations between the parameters and the metrics are complex, we treat the exploration as a black-box optimization problem and formulate it as:

$$y = f(w; x) \qquad \chi = \arg\max_{x \in \mathbb{X}} f(w; x) \qquad (1)$$

where $w$ denotes the target computation workloads, $x$ denotes the accelerator parameters, and $y$ denotes performance metrics. $\mathbb{X}$ is the hardware design space. $f$ is a collection of objective functions that characterize the relationship between the accelerator parameters $x$, workloads $w$, and metrics $y$. $\arg\max$ is to find the parameters $x$ that maximizes $y$. As performance metrics of interests can be multi-dimensional, the problem is a case of multi-objective optimization. Then $\arg\max$ is to find the Pareto optimal set $\chi$ over different metrics.

To solve Equation 1 and find the Pareto set $\chi$ of accelerator parameters, we develop a multi-objective Bayesian optimization (MOBO) algorithm [43], [54], [55] in HASCO. Compared with other black-box optimization methods, Bayesian optimization attempts to find the global optimum in a few steps. It incorporates prior information about the objective function $f$ into a surrogate model, which gives the posterior distribution of $f$. Then Bayesian optimization determines the

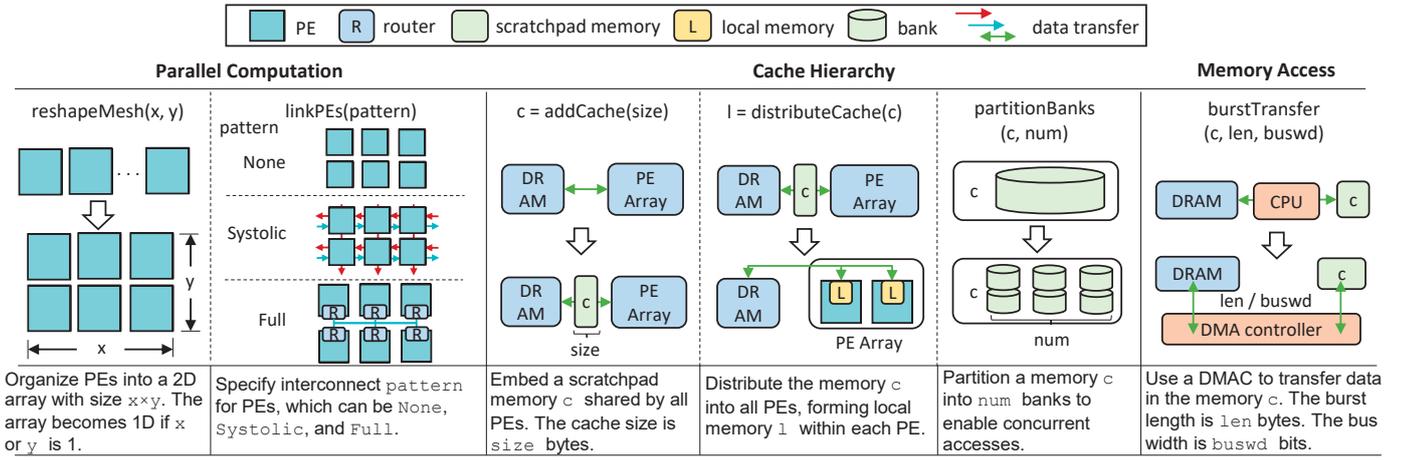

Fig. 6: Main hardware primitives used in HASCO.

**Algorithm 1** Pseudo-code for the MOBO Algorithm

**Input** $\mathbb{X}, f, w, N, \mathbb{M}, ac$
1: Init the prior: $\mathbb{D} \leftarrow \text{sample}(f, \mathbb{X})$
2: **for** $i \leftarrow |\mathbb{D}|$ to $N$ **do**
3:     Update the surrogate model $\mathbb{M}$ to fit $\mathbb{D}$
4:     Calculate the posterior $p(y|x, \mathbb{D})$ with $\mathbb{M}$
5:     Acquire a promising $x_i$:
       $x_i \leftarrow ac(\mathbb{X}, p(y|x, \mathbb{D}))$
6:     Evaluate $x_i$: $y_i \leftarrow f(w; x_i)$
7:     Update the prior: $\mathbb{D} \leftarrow \mathbb{D} \cup (x_i, y_i)$
8:     Calculate the Pareto set: $\chi \leftarrow$ Pareto set of $\mathbb{D}$
9: **end for**
10: Return the current Pareto set $\chi$

most promising $x$ that maximizes $f$ with the posterior and an acquisition function. As the optimization proceeds, the prior information about $f$ and the surrogate model keep updating, resulting in better posterior distributions and $\chi$.

Algorithm 1 gives the overall procedure of the MOBO algorithm. It first samples and evaluates design points to build a training dataset $\mathbb{D}$ incorporating prior information (Line 1). Then it explores the design space iteratively till the maximal trial number $N$ is reached. At each iteration, it updates the surrogate model $\mathbb{M}$ and computes the posterior distribution $p(y|x, \mathbb{D})$ (Line 3-4). Based on the surrogate model, it selects the design point $x_i$ with the acquisition function $ac$ and evaluates $x_i$ (Line 5-6). Last, it updates the prior dataset with the newly explored design point and calculates the Pareto set $\chi$ (Line 7-8). The Pareto set can help us to achieve better trade-offs among different performance constraints in changing scenarios. In practice, we use a Gaussian Process (GP) [64] as the surrogate model and use the hypervolume-based probability of improvement [5] as the acquisition function. The GP model explicitly describes the relation between parameters and metrics is cheap to evaluate.

## VI. SOFTWARE AND INTERFACE GENERATION

Figure 5(a) shows the workflow of the software generation. For each workload, HASCO builds a software design space with software primitives and explores it using heuristic and Q-learning methods. HASCO also generates interfaces dedicated to the target accelerator.

### A. Software Primitives and Design Space

We use a set of software primitives, including partitioning (*tensorize*), reordering (*reorder*), splitting (*split*), fusion (*fuse*), etc. Especially, the *tensorize* primitive uses loops to express a tensorized sub-workload. All the combinations of these primitives form a software design space. Formally, a sequence of software primitives form the skeleton of an optimization, and by setting the factor of each primitive in the sequence, we get a concrete optimization. In Figure 5(c), we show a sequence example of optimizing convolutions for GEMM accelerators. The primitive sequence is [*split*, *reorder*, *fuse*, *tensorize*]. It means we first split the y, k, and c loops into six sub-loops and interchange all loops in a specified order. Then we fuse the four outer-most loops into one loop and specify the three inner-most loops denoted by y2, k2, and c2 as a tensorized sub-workload.

### B. Software Optimization and Generation

Finding the optimal software optimization is an open problem and calls for efficient DSE algorithms. The effect of the optimizations depends on the memory system and compute capability of the target accelerator. The *reorder*, *split*, and *fuse* primitives determine how tensors are accessed in the DRAM and cached off-chip, which in turn affects the software latency. The *tensorize* primitive specifies a sub-workload, and the sub-tensors processed by the sub-workload are all stored in the scratchpad. An optimization is valid only if the actual scratchpad of the target accelerator can accommodate all the sub-tensors. To guarantee the quality of the exploration results, we initialize plenty of candidate optimizations before the exploration starts by randomly generating primitive sequences and factors. Then, we incrementally revise the candidate optimizations to generate new candidates. The revision process may repeat for hundreds of rounds till we find good optimizations. The best optimization would be translated into the final software program by code generation tools [11]. As the exploration proceeds, there can be a great number of candidates in hand, making it time-consuming to revise all of the candidates. Also, to revise each candidate, we have many choices: change the combination of the primitive sequence or change one primitive factor. Exhaustively trying out all the possible revision choices is inefficient.

TABLE I: Benchmark Tensor Computations.

| Tensor Computation | Notation | Workloads | Compute Complexity |
|---|---|---|---|
| MTTKRP | $D[i,j] = \sum A[i,k,l] * B[l,j] * C[k,j]$ | 10 | 255M - 5.9G |
| TTM | $C[i,j,k] = \sum A[i,j,l] * B[l,k]$ | 10 | 16M - 8.6G |
| 2D Conv. | $C[k,x,y] = \sum A[c, x+r, y+s] * B[k,c,r,s]$ | 10 + CNNs | 87M - 3.7G |
| GEMM | $L[i,j] = \sum M[i,k] * N[k,j]$ | 10 | 16K - 4.3G |

We determine what and how to revise in two steps. The first step is to find valuable candidates among all candidate optimizations, and the next step selects the most promising revision choice from all possible choices. There are a number of algorithms for implementing the two steps, such as the random algorithm, dynamic programming, and machine learning algorithms. Especially, the two steps cater to the exploration and exploitation in reinforcement learning [49], [50], [70], [87]. Thus, we use a heuristic algorithm and a Q-learning algorithm to implement the two steps, respectively, as shown in Figure 5(d) and (e). **To identify valuable candidates**, we measure and maintain the latency of each candidate optimization $p$ as $l_p$, and the lowest latency in history is $l^*$. Then, the value of $p$ is measured by $exp(-(l^* - l_p)/l^*)$ [85]. The higher the value is, the better the candidate is. We choose the top-$k$ candidates as valuable candidates, where $k$ is a mutable value. **To revise candidates**, we use Q-learning to generate a new candidate $p'$ for a valuable candidate $p$. In Q-learning, we use a Q-value to indicate how good each revision choice is. We apply the revision choice with the highest Q-value to $p$ to generate $p'$. Specifically, we use the DQN [51] algorithm to train a 4-layer fully-connected neural network, which predicts Q-values. The DQN is reused for all design points in a software space.

### C. Interface Generation

Interfaces and intrinsics are only function abstractions and need to be translated into accelerator instructions. There are two basic types of instructions: the data movement instructions move data between the scratchpad memory and the DRAM, and the compute instructions invoke computations on the PE array. Such ISAs suggest the tensorize interface should explicitly manage scratchpad data and call the intrinsic function. HASCO inserts the data movement instructions before and after the intrinsic call to prepare the scratchpad. Then it replaces the intrinsic call with the compute instructions. For instance, the GEMM intrinsic is replaced with the *compute_accumulated* instruction of Gemmini, which controls the PE array to perform 16 × 16 multiply-add operations.

## VII. EXPERIMENTS

### A. Experimental Setup

**Benchmarks.** We use a set of tensor computations as our benchmarks, as shown in Table I. MTTKRP and TTM are core computations in tensor decomposition. GEMM and 2D convolution are used in convolutional neural networks (CNNs). We also collect workloads from modern CNNs as the 2D convolutions, including ResNet-50 [28], MobileNet [31], and Xception [17].

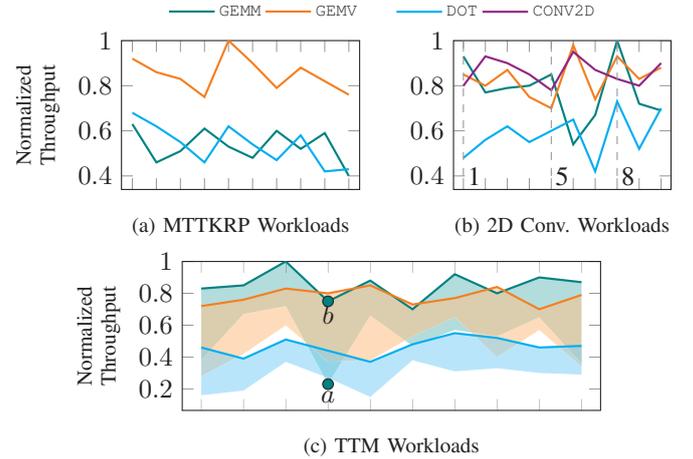

(a) MTTKRP Workloads
(b) 2D Conv. Workloads
(c) TTM Workloads

Fig. 7: Normalized throughput results of different tensor computations and hardware intrinsics.

**Hardware.** In our evaluation, we use four hardware intrinsics: DOT (dot product: $C = \sum A[i] * B[i]$), GEMV ($C[i] = \sum A[i,j] * B[j]$), GEMM, and CONV2D (2D convolution). We employ Gemmini [24] to generate GEMM accelerators. We use the Rocket Chip generator [3] and our Chisel generator [33] to build accelerators with the other intrinsics. For simplicity, we refer to accelerators with GEMM and CONV2D intrinsics as GEMMCore and ConvCore, respectively.

**Methodology.** We first analyze different intrinsics and tensorize choices. Then we demonstrate the efficiency of our hardware DSE with comparisons and detailed analysis. For the software, we compare HASCO with AutoTVM [12] and an accelerator library [24]. The library implements hand-tuned computations, such as matrix multiplication of any size, CNNs, and non-linear activations. It carefully splits and reorders loops in the computations and calls the GEMM intrinsic. Last, we discuss the overall benefits brought by co-design.

**Metrics.** We use Maestro [41], an open-source accelerator microarchitectural model, in the hardware DSE evaluation (Section VII-C). Maestro models spatial accelerators with a scratchpad, local memories, PEs, and interconnections between the PEs. It estimates latency, power, and area by analyzing the reuse across time/space, computations, and memory transactions. However, Maestro omits the modeling of off-chip memory systems. Thus, in the remaining experiments, we synthesize the accelerators with Xilinx Vivado tools [78] and prototype them as Rocket Chip SoCs on a Xilinx VU9P FPGA board. We time the latency, calculate the throughput, and evaluate the chip power with Vivado tools.

### B. Tensorize Choice and Hardware Intrinsic

We compare the four hardware intrinsics (DOT, GEMV, GEMM, and CONV2D) when optimizing the benchmarks'

throughput. Concretely, we specify an array of 64 PEs and a 256 KB scratchpad memory with our hardware primitives for all accelerators and give them different intrinsic functions. With tensor syntax trees and the two-step matching, HASCO can divide all the tensor computations into GEMM, GEMV, and DOT sub-workloads. Only 2D convolutions can be tiled into CONV2D sub-workloads.

Figure 7 compares the throughput results of the four intrinsics. The X-axis represents the workloads of each tensor computation, and the Y-axis is the normalized throughput. We draw two conclusions. First, **different tensor computations prefer different hardware intrinsics**. In general, an intrinsic is more efficient if it is dedicated to the tensor computation. As the figure shows, in most cases, TTM and GEMM prefer the GEMM intrinsic, and 2D convolution prefers the CONV2D intrinsic. Dedicated accelerators provide more data reuse opportunities and achieve higher performance. Although the DOT intrinsic is the most general, it reuses no tensor data within a tensorize interface and achieves low performance. MTTKRP is an exception, which prefers the GEMV intrinsic instead of GEMM. To illustrate the reason clearly, we treat MTTKRP as two stages: $E[i, k, j] = \sum A[i, k, l] * B[l, j]$ and $D[i, j] = \sum E[i, k, j] * C[k, j]$. Only the first $A \times B$ stage can be divided into GEMM sub-workloads and accelerated by the GEMM intrinsic. Nevertheless, HASCO can find GEMV sub-workloads in the two stages from tensor syntax trees. In other words, the GEMM intrinsic accelerates three loops represented by i/k, l, and j in MTTKRP, while the GEMV intrinsic benefits four loops represented by i, k, l, and j.

Even though a hardware intrinsic is dedicated to a tensor computation, the intrinsic does not always achieve the best performance for the computation. Take 2D convolutions, for instance. The product of r and s is termed as the filter size in CNNs. The CONV2D intrinsic processes sub-workloads with a fixed filter size ($3 \times 3$ in our experiments). For a convolution workload, if $r \times s$ is not a multiple of the fixed filter size, the CONV2D intrinsic will conduct redundant computation and become less efficient. In Figure 7(b), the #1, #5, and #8 workloads have $5 \times 5$ and $7 \times 7$ filter sizes, leading to 30.56% and 39.51% redundant computation, respectively. In contrast, the GEMM intrinsic computes sub-workloads in a more fine-grained fashion. By analyzing tensor syntax trees, HASCO determines that three loops of convolutions match the GEMM intrinsic: k, x/y, and c/r/s. Regardless of $r \times s$, the GEMM intrinsic can still exploit the parallelism in the loop c. As a result, the GEMM intrinsic provides the best performance to workloads #1, #5, and #8.

Second, **different tensorize choices have different impacts on the target metrics** (throughput in this case). For each combination of workloads and intrinsics, HASCO can find a great number of tensorize choices and explore them efficiently. In Figure 7(c), we use a colored area to represent the throughput range of the tensorize choices for the same intrinsic. Data reuse, locality, and padding all contribute to the throughput variance. To illustrate the variance clearly, we mark two tensorize choices $a$ and $b$ in the figure. Choice $a$ divides tensor A in TTM along the last two dimensions j and l so that the sub-tensor can be accessed continuously in the DRAM. Choice $b$ divides A along the i and l dimensions, leading to non-continuous data access. Besides, the interface of $a$ calls the GEMM intrinsic exactly 64 times, while the interface of $b$ requires data padding before calling the intrinsic. As a result, the throughput results of the two choices have a 3.26X difference.

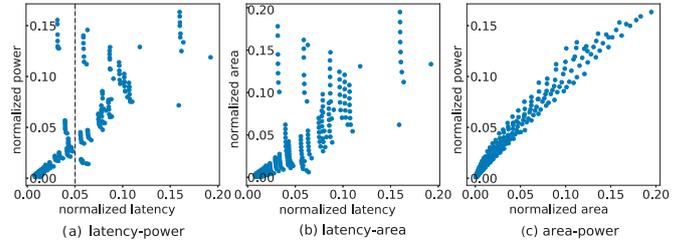

Fig. 8: Correlations between latency, power, and area data collected with Maestro [41].

### C. Hardware DSE Evaluation

**Ground Truth.** We first collect metrics of ConvCore accelerators with models as the ground truth data. We limit the design space of this experiment by simplifying workloads and accelerator parameters. There are six convolutions from Xception ranging from 86.7 MOPs to 454.2 MOPs. We only explore the PE array shape and bank number of the scratchpad memory. All software programs are generated by HASCO.

Figure 8 illustrates the ground truth data. When designing accelerators, the correlations between the latency, power, and area data are non-trivial. Figure 8(c) shows a positive correlation between the normalized power and area data, as a larger design spends more energy on computations and PE communications. However, the normalized power and area can vary dramatically under the same latency constraint, as Figure 8(a) and (b) show. For instance, the power ranges from 207.46 mW to 25136.7 mW under the 0.05 normalized latency constraint, leading to a 121.16X difference. Hence, finding the Pareto solutions to tensor computations is vital to energy-efficient designs.

We further analyze how the accelerator parameters impact the performance metrics. Figure 9 illustrates the correlations between the ground truth data and the parameters. The X-axis represents the bank numbers ranging from one to eight, and the Y-axis represents the PE array shape ranging from $4 \times 4$ to $32 \times 32$. The color of each point indicates a normalized latency, power, or area value. As Figure 9(b) and (c) show, power and area data increase as the PE number and the bank number increase. This observation is natural since the PEs and scratchpad consume more energy and areas. Normally, the latency shows negative correlations with the PE number and the bank number. As the PEs and banks become over-provisioned, the contour color would remain the same. However, in this case, the latency increases when the generated convolution accelerators have more PEs and banks, as Figure 9(a) shows. The reason is that the convolutions used in this ground truth experiment have small computations and limited parallelism. Small PE arrays are enough to process them efficiently. As the PE number keeps growing, the latency of one intrinsic call also increases. Also, more data are padded to fill the PE array, leading to wasted computations and an increase in overall latency.

TABLE II: Pareto solutions of the random search, NSGAII, and MOBO methods. L: latency. P: power.

| Workloads & Constraints | Intrinsic | Latency ($\times 10^8$ cycles) | | | Power (mW) | | | Area ($\times 10^7$ μm$^2$) | | |
|---|---|---|---|---|---|---|---|---|---|---|
| | | Random | NSGAII | MOBO | Random | NSGAII | MOBO | Random | NSGAII | MOBO |
| ResNet: L $\leq$2E9, P$\leq$1E4 | GEMM | 3.972 | 3.528 | 3.528 | 3293.5 | 3099.8 | 3099.8 | 9.470 | 7.005 | 7.005 |
| | CONV2D | 4.439 | 3.863 | 3.411 | 3298.65 | 2989.44 | 2403.87 | 7.469 | 6.551 | 5.438 |
| MobileNet: L $\leq$ 1E10, P$\leq$1E4 | GEMM | 30.62 | 19.23 | 19.23 | 4068.14 | 3874.18 | 3487.17 | 19.33 | 16.86 | 11.93 |
| | CONV2D | 21.35 | 23.35 | 18.13 | 3811.98 | 3589.21 | 3589.21 | 13.04 | 12.29 | 12.29 |
| Xception: L $\leq$1E11, P$\leq$1E4 | GEMM | 228.6 | 228.6 | 228.6 | 4874.48 | 4355.45 | 3874.18 | 27.16 | 24.25 | 16.86 |
| | CONV2D | 237.5 | 217.7 | 217.7 | 4013.59 | 4456.74 | 4013.59 | 17.56 | 19.43 | 17.56 |

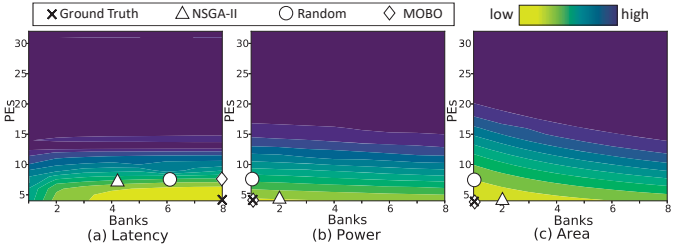

Fig. 9: Correlations between the ground truth data and accelerator parameters.

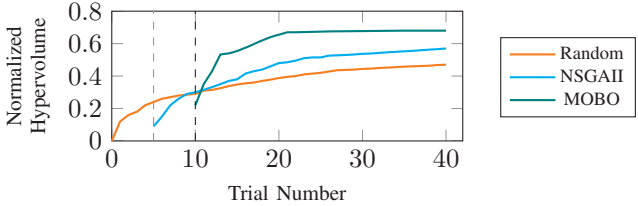

Fig. 10: Normalized hypervolume improvements of random search, NSGAII, and MOBO.

**Comparisons.** We compare the MOBO method used in HASCO with random search and the NSGAII genetic algorithm [22]. We use HASCO to generate software and use the three methods to optimize the latency, power, and area of ConvCore simultaneously within 20 trials[†]. We mark the final solutions in Figure 9. MOBO can find the Pareto optimal set. The random search achieves 1.337X latency, 2.283X power, and 2.404X area compared with the Pareto set. NSGAII achieves 1.242X latency, 1.05X power, and 1.608X area.

We then use more CNNs and intrinsics (GEMM and CONV2D) for comparisons. In the evaluations[†], we constrain the latency and power and use the three methods to find the Pareto solutions. Table II lists the constraints and results. MOBO always outperforms the random search and NSGAII in our evaluations. It achieves 1.215X average latency improvement, 1.154X average power reduction, and 1.336X average area reduction compared with the random search.

We calculate the hypervolume for the case using ResNet and the GEMM intrinsic to show the three methods' convergence. In multi-objective optimizations, the hypervolume indicator measures the size of the space dominated by a set of design points. The closer the design points are to the Pareto front, and the more likely they are distributed along the Pareto front, the larger the hypervolume becomes. As Figure 10 shows, MOBO quickly improves its hypervolume after the initialization phase. It has surpassed the final results for both the NSGAII and random search algorithms at trial 16. The reason is that MOBO reduces the number of redundant evaluations by building a statistical model based on earlier observations. Notably, it takes minutes to hours to model, implement, and profile accelerators per trail. MOBO achieves **a 1.19X hypervolume improvement** compared with NSGAII, meaning MOBO finds more design points close to the Pareto front. It uses **2.5X fewer trials** to achieve the final hypervolume of NSGAII, significantly reducing the co-design cost.

[†]MOBO uses five samples as its prior and iterates 15 times.
[†]The maximal trial number of all methods is set as 40. The population size of NSGAII is 5. The sample size of MOBO is 10.

### D. Software DSE Evaluation

We first demonstrate the software quality by comparing HASCO with the library proposed in [24]. We prototype a GEMMCore on the FPGA, which has a 16 × 16 PE array and a 256 KB scratchpad. Then we use the library to run ResNet on the accelerator. The library converts 2D convolutions to GEMMs and invokes the GEMM intrinsic. Specifically, it always unfolds the operand tensors into matrices (*im2col*), performs GEMMs, and folds the result matrix back to a tensor (*col2im*) [34]. GEMMs converted from convolutions are divided into sub-workloads by loop splitting. The split factors depend on the array shape and scratchpad size.

We use HASCO to optimize software for ResNet and the accelerator. The HASCO-generated software outperforms the library by more than 2X in 18 cases out of ResNet's 53 convolution workloads and provides **a 3.17X average latency reduction**. We illustrate the first 20 cases in Figure 11. As the library converts 2D convolutions to GEMMs, the convolution becomes $C[k, x \times y] = \sum A[c \times r \times s, x \times y] * B[k, c \times r \times s]$. This conversion can be omitted only if the r and s loops are reduced. It is an algorithm-level optimization beyond the scope of this paper. Though the conversion is a natural way to call GEMM intrinsics, it introduces significant latency overheads. As Figure 11 shows, once the *im2col* and *col2im* are performed, their overhead dominates the overall latency of the workload. Additionally, the conversion requires a much larger DRAM region to store the intermediate matrices. In contrast, HASCO customizes tensorize interfaces for each GEMM workload. Instead of converting convolutions to GEMMs, it directly partitions a convolution workload along different dimensions according to operand tensors' shapes. For convolutions where dimension c is large, HASCO would choose c as a partition dimension to provide enough data parallelism. Otherwise, dimension r/s would be partitioned to reduce wasted computations as much as possible.

TABLE III: HASCO results when scaling the power constraints. Mem: Memory. Bk: Banks.

| Scenario | CNNs | Baseline-GEMMCore: separated | | | | HASCO-GEMMCore: co-design | | | | HASCO-ConvCore: co-design | | | | HLS-Core | |
|---|---|---|---|---|---|---|---|---|---|---|---|---|---|---|---|
| | | PEs | Mem (KB) | Bk | latency (ms) | PEs | Mem (KB) | Bk | latency (ms) | PEs | Mem (KB) | Bk | latency (ms) | PEs | latency (ms) |
| Edge (power: 2 W) | ResNet | 64 | 256 | 4 | 12321.8 | 64 | 256 | 6 | 8547.8 | 144 | 320 | 8 | 4673.7 | 144 | 8931.7 |
| | MobileNet | 64 | 256 | 4 | 56457.6 | 64 | 512 | 8 | 42977.1 | 121 | 512 | 6 | 24273.5 | 121 | 49828.9 |
| | Xception | 64 | 256 | 4 | 707105.1 | 64 | 512 | 8 | 544023.8 | 144 | 512 | 8 | 318485.6 | 144 | 693601.4 |
| Cloud (power: 20 W) | ResNet | 4096 | 1024 | 4 | 260.5 | 4096 | 1024 | 8 | 197.2 | 4096 | 1536 | 8 | 195.3 | 4096 | 315.4 |
| | MobileNet | 4096 | 1024 | 4 | 1456.9 | 4096 | 1024 | 8 | 1020.5 | 4096 | 1024 | 8 | 901.5 | 4096 | 1580.3 |
| | Xception | 4096 | 1024 | 4 | 15706.3 | 4096 | 1024 | 8 | 12548.9 | 4096 | 1536 | 8 | 11594.4 | 4096 | 22189.7 |

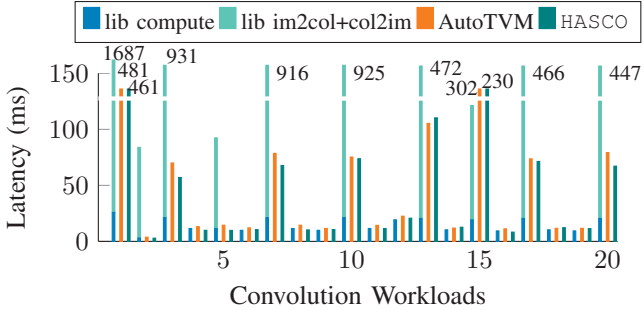

Fig. 11: Comparisons between ResNet software.

For a fair comparison, we also use AutoTVM [12] to optimize software by directly partitioning convolutions. AutoTVM requires users to manually make tensorize choices and write primitive templates for each tensor computation. Besides, it only optimizes the size of tensorized sub-workloads. For ResNet and the GEMMCore, **HASCO outperforms AutoTVM by 1.21X**. The improvement is because HASCO systematically explores tensorize choices and software primitives, while AutoTVM relies on static templates and fixed tensorize choices.

### E. Overall Solution Analysis

Last, we evaluate our hardware and software DSE algorithms together and demonstrate the overall benefits brought by co-design. We scale the power constraints to simulate cloud (20 W) and edge (2 W) scenarios, respectively. Under the constraints, we use HASCO to generate GEMMCore accelerators and software in 20 co-design iterations. Table III lists the accelerator parameters and latency results.

The baseline employs the traditional methodology, which decouples the hardware and software developments. For the baseline hardware, we employ two GEMMCore accelerators with the default parameters listed in Table III. If we use the library [24] as our baseline software, HASCO can achieve a 2.14X average latency reduction. For fair comparisons and improvement breakdowns, we use AutoTVM to generate the baseline software. As the table shows, HASCO solutions achieve **a 1.25X to 1.44X latency reduction** in the two scenarios compared with the baseline solutions. The hardware DSE of HASCO provides 29.53% of the latency reduction. As the table shows, the GEMMCore accelerators generated by HASCO tend to use more scratchpad memories and banks, which enable larger tensorized sub-workloads and more data reuse. For each scenario, the HASCO accelerator uses the same number of PEs as the baseline. The reason is GEMMCore constrains its PE array shape to be $2^n \times 2^n$. Under this PE constraint and the power constraint, MOBO converges to the optimal PE array shape. The software optimization of HASCO provides the remaining latency reduction.

We then use HASCO to co-design ConvCore accelerators and the software. The results are also given in Table III. The HASCO-ConvCore solutions further reduce the latency by 1.42X on average compared with the HASCO-GEMMCore solutions. The improvement is two-fold. For one thing, the CONV2D intrinsic is dedicated to convolutions and provides more data reuse opportunities than the GEMM intrinsic. For another, unlike GEMMCore, ConvCore does not restrict the PE array shape, giving HASCO opportunities to use more PEs under the same power constraint.

We also compare with HLS-based solutions for convolution accelerator by using Vivado tools. For an HLS-solution, all the workloads are synthesized into one hardware, and we refer to the hardware as HLS-Core. In our implementation of HLS-Cores, we unroll the c and k loops to provide sufficient parallelism and synthesize the remaining loops into datapaths. As Table III (Column HLS-Core) shows, ConvCores achieve 1.615X to 2.181X latency improvements compared with HLS-Cores. The reason is convolutions in CNNs differ in tensor sizes and require specialized loop optimization. The datapaths in HLS-Cores lead to fixed sub-workload sizes and loop orders, making HLS-Cores only efficient for a small portion of convolutions. In contrast, HASCO generates an optimized software for each workload. By orchestrating loop orders and split factors in the software, HASCO can exploit parallelism in different loops and form sub-workloads in different sizes. In this way, the software provides flexibility to ConvCores to handle different convolutions efficiently. Moreover, for a complicated application using multiple tensor computations [38], HLS needs to generate hardware for each computation, while HASCO only generates one hardware.

## VIII. RELATED WORKS

**Hardware Acceleration**. Many hardware accelerators have been proposed for DNNs and tensor computations. Previous works [13]–[15], [23], [26], [27], [35], [44], [45], [62], [81]–[84], [86] target the most common computations in DNNs, including convolutions and matrix multiplications. Previous works [25], [29], [42], [57], [63], [69] propose flexible archi-tectures for more general tensor computations. All these works generate chips with fixed powers and areas and cannot be scaled across cloud and edge devices. Recently, hardware generators are proposed to provide more efficiency. Gemmini [24] generates systolic array accelerators for matrix multiplications. NVDLA [21] generates deep learning inference accelerators scaled across a wide range of IoT devices. MAGNet [74] and AutoDNNchip [79] are generator infrastructures generating DNN accelerators. MAGNet is based on highly configurable PEs and supports multiple dataflows. AutoDNNchip instantiates IPs to generate accelerators rapidly. DSAGEN [76] extracts information about parallelism and concurrency from the target workload and generates specialized spatial accelerators from scratch. VTA [52] designs parameterizable architectures, where memories, datatypes, and sizes of the GEMM intrinsic can be customized. It relies on AutoTVM [12] to optimize software, meaning users make tensorize choices. HASCO exploits off-the-shelf hardware generators.

**Design Space Exploration**. For software optimizations, loop transformations [8], [9], [48], [61], [75], [77] have been studied for decades. The traditional flows use heuristic algorithms to optimize software. Recently, loop transformations using machine learning algorithms have been proposed. Halide auto-scheduler [1] uses tree searching and random programs in the exploration and mainly targets image processing. PlaidML [18] and Tensor Comprehensions [73] use an analytical model and polyhedral models for software DSE, respectively. Halide, PlaidML, and Tensor Comprehensions only support limited heterogeneous hardware platforms. AutoTVM [12] uses XGBoost [10] for exploration and supports more platforms. It requires programmers to develop primitive templates and make tensorize choices. FlexTensor [85] proposes a fully-automatic method to optimize programs. However, it only supports general programming platforms. HASCO targets spatial accelerators with different intrinsics.

Many hardware generators design DSE methods for their accelerators, such as [74], [76], [79], [80]. For instance, AutoDNNchip [79] also builds models to predict accelerator metrics based on DNN parameters. It relies on design space pruning to enable fast exploration. DSAGEN [76] iteratively optimizes a single objective until the objective converges. ConfuciuX [37] uses reinforcement learning and genetic algorithms to search the number of PEs assigned to different DNN layers. It leverages accelerators in a fine-grained way, which requires architecture supports. Besides, it optimizes one objective at a time. Compared with these works, HASCO provides a multi-objective DSE approach serving a class of spatial accelerators. Interstellar [80] analyzes the impact of blocking and different dataflows on the DNN accelerators. It transforms loops to fit into the resource-constrained hardware but lacks systematic software optimization. More importantly, Interstellar cannot reuse the generated accelerators for other tensor computations as it is unaware of hardware intrinsics. In contrast, HASCO automatically explores numerous tensorize choices and jointly optimizes the hardware and software. The software retains flexibility such that the hardware serves multiple tensor computations.

## IX. CONCLUSION

Though HW/SW co-design can generate high-quality solutions to tensor computation, it faces two fundamental challenges: identifying substantial partitioning methods and efficiently exploring the huge hardware-software design space. In this work, we propose HASCO as an agile co-design approach. HASCO automatically identifies partitioning methods from tensor syntax trees. It uses heuristic and Q-learning algorithms for software optimization. It uses multi-objective Bayesian optimization to explore hardware parameters. Putting these techniques together, HASCO provides significant improvements in solution quality and DSE efficiency.

ACKNOWLEDGMENT

This work is supported by the Beijing Natural Science Foundation (No. JQ19014), Beijing Academy of Artificial Intelligence (BAAI), and Key-Area Research and Development Program of Guangdong Province (No. 2019B010155002).